\documentstyle[11pt]{article}
\begin{document}

\title{{Interacting new agegraphic tachyon, K-essence and dilaton scalar field models of dark energy in non-flat universe}}
\author{K. Karami\thanks{E-mail: KKarami@uok.ac.ir}\\M.S. Khaledian\\F. Felegary\\Z. Azarmi\\
\small{Department of Physics, University of Kurdistan, Pasdaran
St., Sanandaj, Iran}\\
}

\maketitle

\begin{abstract}
We study the correspondence between the tachyon, K-essence and
dilaton scalar field models with the interacting new agegraphic
dark energy model in the non-flat FRW universe. We reconstruct the
potentials and the dynamics for these scalar field models, which
describe accelerated expansion of the universe.
\end{abstract}
\clearpage
\section{Introduction}
Type Ia supernovae observational data suggest that the universe is
dominated by two dark components containing dark matter and dark
energy \cite{Riess}. Dark matter (DM), a matter without pressure,
is mainly used to explain galactic curves and large-scale
structure formation, while dark energy (DE), an exotic energy with
negative pressure, is used to explain the present cosmic
accelerating expansion. However, the nature of DE is still
unknown, and people
 have proposed some candidates to describe it. The cosmological
 constant, $\Lambda$, is the most obvious theoretical candidate of
 DE, which has the equation of state $\omega=-1$.
 Astronomical observations indicate that the cosmological constant is many orders of magnitude
 smaller than estimated in modern theories of elementary particles \cite{Weinberg}. Also the
 "fine-tuning" and the "cosmic coincidence" problems are the two
 well-known difficulties of the cosmological constant problems
 \cite{Copeland}.

 There are different alternative theories for the dynamical DE scenario which have been
 proposed by people to interpret the accelerating universe. i) The scalar-field models of DE including
 quintessence \cite{Wetterich},
 phantom (ghost) field \cite{Caldwell1}, K-essence \cite{Chiba}
 based on earlier work of K-inflation \cite{Picon3},
 tachyon field \cite{Sen}, dilatonic ghost condensate \cite{Gasperini},
 quintom \cite{Elizalde}, and so forth. ii) The interacting DE models including
 Chaplygin gas \cite{Kamenshchik}, holographic DE models
 \cite{Cohen}, and braneworld models
 \cite{Deffayet}, etc.

Recently, the original agegraphic dark energy (OADE) and new
agegraphic dark energy (NADE) models were proposed by Cai
\cite{Cai} and Wei $\&$ Cai \cite {Wei1}, respectively. Cai
\cite{Cai} proposed the OADE model to explain the accelerated
expansion of the universe, based on the uncertainty relation of
quantum mechanics as well as the gravitational effect in general
relativity. The OADE model had some difficulties. In particular,
it cannot justify the matter-dominated era \cite{Cai}. This
motivated Wei and Cai \cite{Wei1} to propose the NADE model, while
the time scale is chosen to be the conformal time instead of the
age of the universe. The evolution behavior of the NADE is very
different from that of the OADE. Instead the evolution behavior of
the NADE is similar to that of the holographic DE \cite{Cohen}.
But some essential differences exist between them. In particular,
the NADE model is free of the drawback concerning causality
problem which exists in the holographic DE model. The ADE models
assume that the observed DE comes from the spacetime and matter
field fluctuations in the universe \cite{Wei1,Wei2}. The ADE
models have been studied in ample detail by
\cite{Kim,Sheykhi,Wei3}.

Besides, as usually believed, an early inflation era leads to a
flat universe. This is not a necessary consequence if the number
of e-foldings is not very large \cite{Huang}. It is still possible
that there is a contribution to the Friedmann equation from the
spatial curvature when studying the late universe, though much
smaller than other energy components according to observations.
Therefore, it is not just of academic interest to study a universe
with a spatial curvature marginally allowed by the inflation model
as well as observations. Some experimental data have implied that
our universe is not a perfectly flat universe and that it
possesses a small positive curvature \cite{Bennett}.

All mentioned in above motive us to consider the interacting NADE
model in the non-flat FRW and investigate its correspondence with
the tachyon, K-essence and dilaton scalar field models. To do
this, in Section 2, we obtain the equation of state parameter for
the NADE model in a non-flat universe. In Section 3, we suggest a
correspondence between the NADE and the tachyon, K-essence and
dilaton scalar field models in the presence of a spatial
curvature. We reconstruct the potentials and the dynamics for
these scalar field models, which describe accelerated expansion.
Section 4 is devoted to conclusions.
\section{Interacting NADE model in a non-flat universe}

We consider the Friedmann-Robertson-Walker (FRW) metric for the
non-flat universe as
\begin{equation}
{\rm d}s^2=-{\rm d}t^2+a^2(t)\left(\frac{{\rm
d}r^2}{1-kr^2}+r^2{\rm d}\Omega^2\right),\label{metric}
\end{equation}
where $k=0,1,-1$ represent a flat, closed and open FRW universe,
respectively. Observational evidences support the existence of a
closed universe with a small positive curvature ($\Omega_{k}\sim
0.02$) \cite{Bennett}.

For the non-flat FRW universe containing the DE and DM, the first
Friedmann equation has the following form
\begin{equation}
{\textsl{H}}^2+\frac{k}{a^2}=\frac{1}{3M_p^2}~
(\rho_{\Lambda}+\rho_{\rm m}),\label{eqfr}
\end{equation}
where $\rho_{\Lambda}$ and $\rho_{\rm m}$ are the energy density
of DE and DM, respectively. Let us define the dimensionless energy
densities as
\begin{equation}
\Omega_{\rm m}=\frac{\rho_{\rm m}}{\rho_{\rm cr}}=\frac{\rho_{\rm
m}}{3M_p^2H^2},~~~~~~\Omega_{\rm
\Lambda}=\frac{\rho_{\Lambda}}{\rho_{\rm
cr}}=\frac{\rho_{\Lambda}}{3M_p^2H^2},~~~~~~\Omega_{k}=\frac{k}{a^2H^2},
\label{eqomega}
\end{equation}
then, the first Friedmann equation yields
\begin{equation}
\Omega_{\rm m}+\Omega_{\Lambda}=1+\Omega_{k}.\label{eq10}
\end{equation}
Following \cite{Sheykhi}, the energy density of the NADE is given
by
\begin{equation}
\rho_{\Lambda}=\frac{3{n}^2M_p^2}{\eta^2},\label{NADE}
\end{equation}
where the numerical factor 3$n^2$ is introduced to parameterize
some uncertainties, such as the species of quantum fields in the
universe, the effect of curved spacetime (since the energy density
is derived for Minkowski spacetime), and so on. The astronomical
data for the NADE gives the best-fit value (with 1$\sigma$
uncertainty) $n = 2.716_{-0.109}^{+0.111}$ \cite{Wei3}. Also
$\eta$ is conformal time of the FRW universe, and given by
\begin{equation}
\eta=\int\frac{{\rm d}t}{a}=\int_{0}^{a}\frac{{\rm d}a}{Ha^2}.
\end{equation}

From definition $\rho_{\Lambda}=3M_p^2H^2\Omega_{\Lambda}$, we get
\begin{equation}
\eta=\frac{{n}}{H\sqrt{\Omega_{\Lambda}}}.\label{eta}
\end{equation}

We consider a universe containing an interacting NADE density
$\rho_{\Lambda}$ and the cold dark matter (CDM), with $\omega_{\rm
m}=0$. The energy equations for NADE and CDM are
\begin{equation}
\dot{\rho}_{\Lambda}+3H(1+\omega_{\Lambda})\rho_{\Lambda}=-Q,\label{eqpol}
\end{equation}
\begin{equation}
\dot{\rho}_{\rm m}+3H\rho_{\rm m}=Q,\label{eqCDM}
\end{equation}
where following \cite{Kim06}, we choose $Q=\Gamma\rho_{\Lambda}$
as an interaction term and
$\Gamma=3b^2H(\frac{1+\Omega_{k}}{\Omega_{\Lambda}})$ is the decay
rate of the NADE component into CDM with a coupling constant
$b^2$. Although this expression for the interaction term may look
purely phenomenological but different Lagrangians have been
proposed in support of it \cite{Tsujikawa}. The choice of the
interaction between both components was to get a scaling solution
to the coincidence problem such that the universe approaches a
stationary stage in which the ratio of DE and DM becomes a
constant \cite{Hu}. Note that choosing the $H$ in the $Q$-term is
motivated purely by mathematical simplicity. Because from the
continuity equations, the interaction term should be proportional
to a quantity with units of inverse of time. For the latter the
obvious choice is the Hubble factor $H$. The dynamics of
interacting DE models with different $Q$-classes have been studied
in ample detail by \cite{Amendola}. It should be emphasized that
this phenomenological description has proven viable when
contrasted with observations, i.e., SNIa, CMB, large scale
structure, $H(z)$, and age constraints \cite{Wang8}, and recently
in galaxy clusters \cite{Bertolami8}.

Taking the time derivative of Eq. (\ref{NADE}), using
$\dot{\eta}=1/a$ and Eq. (\ref{eta}) yields
\begin{equation}
\dot{\rho}_{\Lambda}=-\frac{2H\sqrt{\Omega_{\Lambda}}}{{
n}a}\rho_{\Lambda}.\label{rhodot}
\end{equation}
Substituting Eq. (\ref{rhodot}) in (\ref{eqpol}), gives the
equation of state (EoS) parameter of the interacting NADE model as
\begin{eqnarray}
\omega_{\Lambda}=-1+\frac{2\sqrt{\Omega_{\Lambda}}}{3{
n}a}-b^2\Big(\frac{1+\Omega_{k}}{\Omega_{\Lambda}}\Big).\label{w}
\end{eqnarray}
\section{Correspondence between the interacting NADE and scalar field models of DE}
Here we suggest a correspondence between the interacting NADE
model with the tachyon, K-essence and dilaton scalar field models
in the non-flat universe. To establish this correspondence, we
compare the interacting NADE density (\ref{NADE}) with the
corresponding scalar field model density and also equate the
equations of state for this models with the EoS parameter given by
(\ref{w}).

\subsection{New agegraphic tachyon model} The tachyon field was
proposed as a source of the dark energy. The tachyon is an
unstable field which has become important in string theory through
its role in the Dirac-Born–-Infeld (DBI) action which is used to
describe the D-brane action \cite{Sen}. The effective Lagrangian
density of tachyon matter is given by \cite{Sen}
\begin{equation}
{\mathcal{L}}=-V(\phi)\sqrt{1+\partial_{\mu}\phi
\partial^{\mu}\phi}.
\end{equation}
The energy density and pressure for the tachyon field are as
following \cite{Sen}
\begin{equation}
\rho_{T}=\frac{V(\phi)}{\sqrt{1-\dot{\phi}^{2}}},\label{rhot}
\end{equation}
\begin{equation}
p_{T}=-V(\phi)\sqrt{1-\dot{\phi}^{2}},
\end{equation}
where $V(\phi)$ is the tachyon potential. The EoS parameter for
the tachyon scalar field is obtained as
\begin{equation}
\omega_{T}=\frac{p_{T}}{\rho_{T}}=\dot{\phi}^{2}-1.\label{wt}
\end{equation}
If we establish the correspondence between the NADE and tachyon
DE, then using Eqs. (\ref{w}) and (\ref{wt}) we have
\begin{equation}
\omega_\Lambda=-1+\frac{2\sqrt{\Omega_\Lambda}}{3na}-b^2\Big(\frac{1+\Omega_k}{\Omega_\Lambda}\Big)=\dot{\phi}^{2}-1\label{aget1},
\end{equation}
also comparing Eqs. (\ref{NADE}) and (\ref{rhot}) one can write
\begin{equation}
\rho_{\Lambda}=\frac{3{n}^2M_p^2}{\eta^2}=\frac{V(\phi)}{\sqrt{1-\dot{\phi}^{2}}}\label{aget}~.
\end{equation}
From Eqs. (\ref{aget1}) and (\ref{aget}), one can obtain the
kinetic energy term and the tachyon potential energy as follows
\begin{equation}
\dot{\phi}^{2}=\frac{2\sqrt{\Omega_\Lambda}}{3na}-b^2\Big(\frac{1+\Omega_k}{\Omega_\Lambda}\Big)\label{phi2},
\end{equation}
\begin{equation}
V(\phi)=3M_p^2H^2\Omega_\Lambda\sqrt{1-\frac{2\sqrt{\Omega_\Lambda}}{3na}+b^2\Big(\frac{1+\Omega_k}{\Omega_\Lambda}\Big)}.
\end{equation}
From (\ref{phi2}), we obtain the evolutionary form of the tachyon
scalar field as
\begin{equation}
\phi(a)-\phi(1)=\int_{1}^{a}{\frac{{\rm
d}a}{Ha}\sqrt{\frac{2\sqrt{\Omega_\Lambda}}{3na}-b^2\Big(\frac{1+\Omega_k}{\Omega_\Lambda}\Big)}}\label{phi3},
\end{equation}
where we take $a_0=1$ for the present time. The above integral
cannot be taken analytically. But for the late-time universe, i.e.
$\Omega_\Lambda=1$, $\Omega_k=0$ and $b=0$, Eq. (\ref{phi3})
reduces to
\begin{equation}
\phi(a)-\phi(1)=\int_{1}^{a}{\frac{{\rm
d}a}{Ha}\sqrt{\frac{2}{3na}}}\label{phi33},
\end{equation}
and the Hubble parameter from Eqs. (\ref{eqfr}) and (\ref{rhodot})
can be obtained as
\begin{equation}
H = H_0  \exp{\Big[\frac{1}{n}\Big(\frac{1}{a} -
1\Big)\Big]}\label{H},
\end{equation}
where $H_0$ is the Hubble parameter at the present time. Finally,
using Eqs. (\ref{phi33}) and (\ref{H}) one can get
\begin{equation}
\phi(a)-\phi(1)= \frac{{2e^{1/n} }}{H_0}\sqrt {\frac{{2\pi }}{3}}
~\Big[ {\rm erf}\Big(\frac{1}{{\sqrt n }}\Big) - {\rm
erf}\Big(\frac{1}{{\sqrt {na} }}\Big)\Big],
\end{equation}
where ${\rm erf}(x)$ is the error function defined as
\begin{equation}
{\rm erf}(x) = \frac{2}{{\sqrt \pi }}\int_0^x {e^{ - t^2 } {\rm d}
t}.
\end{equation}
\subsection{New agegraphic K-essence model}
The K-essence scalar field model of DE is also used to explain the
observed late-time acceleration of the universe. The K-essence is
described by a general scalar field action which is a function of
$\phi$ and $\chi=\dot{\phi}^2/2$, and is given by \cite{Chiba,
Picon3}
\begin{equation}
S=\int {\rm d} ^{4}x\sqrt{-{\rm g}}~p(\phi,\chi),
\end{equation}
where $p(\phi,\chi)$ corresponds to a pressure density as
\begin{equation}
p(\phi,\chi)=f(\phi)(-\chi+\chi^{2}),
\end{equation}
and the energy density of the field $\phi$ is
\begin{equation}
\rho(\phi,\chi)=f(\phi)(-\chi+3\chi^{2}).\label{rhok}
\end{equation}
The EoS parameter for the K-essence scalar field is obtained as
\begin{equation}
\omega_{K}=\frac{p(\phi,\chi)}{\rho(\phi,\chi)}=\frac{\chi-1}{3\chi-1}.\label{wk}
\end{equation}
Equating Eq. (\ref{wk}) with the new agegraphic EoS parameter
(\ref{w}), $\omega_{K}=\omega_{\Lambda}$, we find the solution
for $\chi$
\begin{equation}
\chi=\frac{2-\frac{2\sqrt{\Omega_{\Lambda}}}{3{
n}a}+b^2\Big(\frac{1+\Omega_{k}}{\Omega_{\Lambda}}\Big)}{4-\frac{2\sqrt{\Omega_{\Lambda}}}{
na}+3b^2\Big(\frac{1+\Omega_{k}}{\Omega_{\Lambda}}\Big)}\label{chi}.
\end{equation}
Using Eq. (\ref{chi}) and $\dot{\phi}^2=2\chi$, we obtain the
evolutionary form of the K-essence scalar field as
\begin{equation}
\phi(a)-\phi(1)=\int_1^a\frac{{\rm
d}a}{Ha}\left(\frac{4-\frac{4\sqrt{\Omega_{\Lambda}}}{3{
n}a}+2b^2\Big(\frac{1+\Omega_{k}}{\Omega_{\Lambda}}\Big)}{4-\frac{2\sqrt{\Omega_{\Lambda}}}
{na}+3b^2\Big(\frac{1+\Omega_{k}}{\Omega_{\Lambda}}\Big)}\right)^{1/2}\label{phik}.
\end{equation}
For the late-time universe, i.e. $\Omega_\Lambda=1$, Eq.
(\ref{phik}) by the help of Eq. (\ref{H}) yields
\begin{equation}
\phi (a) - \phi (1) = \frac{e^{1/n}}{H_0}\int_1^a
\frac{e^{-1/na}}{a}
 \sqrt {\frac{{1 - 3na}}{{1 -
2na}}}{~\rm d}a\label{phik1}.
\end{equation}
Expanding the integrand appeared in Eq. (\ref{phik1}) for $a>1$,
one can obtain
\begin{equation}
\phi(a)-\phi(1)=\frac{{e^{1/n}}}{{H_0}}\Big[\ln
a+\frac{11}{12n}\Big(\frac{1}{a}-1\Big)-\frac{131}{576n^2}\Big(\frac{1}{a^2}-1\Big)+
\frac{503}{10368n^3}\Big(\frac{1}{a^3}-1\Big)+O\Big(\frac{1}{a^4}-1\Big)\Big].
\end{equation}
\subsection{New agegraphic dilaton field}
The dilaton scalar field model of DE is obtained from the low-energy
limit of string theory. It is described by a general
four-dimensional effective low-energy string action. The coefficient
of the kinematic term of the dilaton can be negative in the Einstein
frame, which means that the dilaton behaves as a phantom-type scalar
field. However, in presence of higher-order derivative terms for the
dilaton field $\phi$ the stability of the system is satisfied even
when the coefficient of $\dot{\phi}^2$ is negative \cite{Gasperini}.
The pressure (Lagrangian) density and the energy density of the
dilaton DE model is given by \cite{Gasperini}
\begin{equation}
p_{D}=-\chi+ce^{\lambda\phi}\chi^{2},
\end{equation}
\begin{equation}
\rho_{D}=-\chi+3ce^{\lambda\phi}\chi^{2},\label{rhod}
\end{equation}
where $c$ and $\lambda$ are positive constants and
$\chi=\dot{\phi}^2/2$. The EoS parameter for the dilaton scalar
field is given by
\begin{equation}
\omega_{D}=\frac{p_D}{\rho_D}=\frac{-1+ce^{\lambda\phi}\chi}{-1+3ce^{\lambda\phi}\chi}.\label{wd}
\end{equation}
Equating Eq. (\ref{wd}) with the new agegraphic EoS parameter
(\ref{w}), $\omega_{D}=\omega_{\Lambda}$, we find the following
solution
\begin{equation}
ce^{\lambda\phi}\chi=\frac{2-\frac{2\sqrt{\Omega_\Lambda}}{3na}+b^2(\frac{1+\Omega_k}{\Omega_\Lambda})}
{4-\frac{2\sqrt{\Omega_\Lambda}}{na}+3b^2(\frac{1+\Omega_k}{\Omega_\Lambda})},
\end{equation}
then using $\chi=\dot{\phi}^2/2$, we obtain
\begin{equation}
e^{\frac{\lambda\phi}{2}}\dot{\phi}=\left(\frac{4-\frac{4\sqrt{\Omega_\Lambda}}{3na}+2b^2(\frac{1+\Omega_k}{\Omega_\Lambda})}
{c\Big(4-\frac{2\sqrt{\Omega_\Lambda}}{na}+3b^2(\frac{1+\Omega_k}{\Omega_\Lambda})\Big)}\right)^{1/2}.
\end{equation}
Integrating with respect to $a$ we get
\begin{equation}
e^\frac{\lambda\phi(a)}{2}=e^\frac{\lambda\phi(1)}{2}+\frac{\lambda}{2\sqrt{c}}\int_1^a\frac{{\rm
d}a}{Ha}\left(\frac{4na\Omega_\Lambda-\frac{4}{3}\Omega_\Lambda^{\frac{3}{2}}+2nab^2(1+\Omega_k)}
{4na\Omega_\Lambda-2\Omega_\Lambda^{\frac{3}{2}}+3nab^2(1+\Omega_k)}\right)^{1/2}\label{phiD}.
\end{equation}
For the late-time universe, Eq. (\ref{phiD}) by the help of Eq.
(\ref{H}) yields
\begin{equation}
e^\frac{\lambda\phi(a)}{2} =e^\frac{\lambda\phi(1)}{2}+
\frac{\lambda e^{1/n}}{2H_0\sqrt{c}}\int_1^a \frac{e^{-1/na}}{a}
 \sqrt {\frac{{1 - 3na}}{{1 -
2na}}}~{\rm d}a.\label{phiD1}
\end{equation}
Expanding the integrand appeared in Eq. (\ref{phiD1}) for $a>1$,
one can obtain
\begin{eqnarray}
\phi(a)=\frac{2}{\lambda}\ln\left[e^\frac{\lambda\phi(1)}{2}+\frac{{\lambda
e^{1/n}}}{{2H_0}{\sqrt{c}}}\Big[\ln
a+\frac{11}{12n}\Big(\frac{1}{a}-1\Big)-\frac{131}{576n^2}\Big(\frac{1}{a^2}-1\Big)\right.\nonumber\\\left.+\frac{503}{10368n^3}\Big(\frac{1}{a^3}-1\Big)
+O\Big(\frac{1}{a^4}-1\Big)\Big]\right].
\end{eqnarray}
\section{Conclusions}

Here we considered the interacting NADE model with CDM in the
non-flat FRW universe. We established a correspondence between the
NADE density with the tachyon, K-essence and dilaton energy
density in the non-flat FRW universe. We reconstructed the
potentials and the dynamics of these scalar field models which
describe tachyon, K-essence and dilaton cosmology.

\end{document}